\documentclass[letterpaper,twocolumn,10pt]{article}
\usepackage{usenix2019_v3}

\usepackage{tikz}
\usepackage{amsmath}

\usepackage{amsfonts}
\usepackage{graphicx}
\usepackage{textcomp}
\usepackage{xcolor}
\usepackage[many]{tcolorbox}
\usepackage{booktabs}
\usepackage{multirow}
\usepackage{array}

\newcolumntype{C}[1]{>{\centering\arraybackslash}p{#1}}

\usepackage{newtxtext}
\usepackage{newtxmath}
\usepackage{algorithm}
\usepackage{algpseudocode}

\usepackage{xspace}
\newcommand{\system}{{\sc ADD}\xspace}
\newcommand{\paragraphbe}[1]{\smallskip\noindent{\bf {#1}.}~}

\definecolor{grey}{HTML}{d3d3d3}
\newtcolorbox{boxA}{
    boxrule = 1.5pt,
    rounded corners,
    colback = grey, 
    boxsep=0.1pt,
}

\begin{document}

\date{}

\title{\Large \bf Defending against Adversarial Malware Attacks on ML-based Android Malware Detection Systems}

\author{
\rm Ping He$^{\dag, \S}$, \xspace \rm Lorenzo Cavallaro$^{\S}$, \xspace \rm Shouling Ji$^{\dag}$\\
$^{\dag}$ {\rm Zhejiang University}, $^{\S}$ {\rm University College London}\\
\rm E-mails: \xspace {\rm gnip@zju.edu.cn}, \xspace {\rm l.cavallaro@ucl.ac.uk}, \xspace {\rm sji@zju.edu.cn}} 

\maketitle

\begin{abstract}
Android malware presents a persistent threat to users' privacy and data integrity.
To combat this, researchers have proposed machine learning-based (ML-based) Android malware detection (AMD) systems.
However, adversarial Android malware attacks compromise the detection integrity of the ML-based AMD systems, raising significant concerns.
Existing defenses against adversarial Android malware provide protections against feature space attacks which generate adversarial feature vectors only, leaving protection against realistic threats from problem space attacks which generate real adversarial malware an open problem.
In this paper, we address this gap by proposing \system, a practical adversarial Android malware defense framework designed as a plug-in to enhance the adversarial robustness of the ML-based AMD systems against problem space attacks.
Our extensive evaluation across various ML-based AMD systems demonstrates that \system is effective against state-of-the-art problem space adversarial Android malware attacks.
Additionally, \system shows the defense effectiveness in enhancing the adversarial robustness of real-world antivirus solutions.
\end{abstract}

\section{Introduction}

As the Android operating system powers a vast array of mobile devices worldwide, it has become a significant target for malware.
Recent security reports have identified over 35 million Android malware samples since 2008, posing substantial threats to users' privacy and data integrity~\cite{MalwareSample}.
To counteract these threats, machine learning-based (ML-based) Android malware detection (AMD) systems have been developed and widely deployed in real-world mobile devices, offering powerful defense effectiveness against Android malware.
Unfortunately, these ML-based AMD systems are vulnerable to problem space adversarial malware attacks~\cite{DBLP:conf/sp/PierazziPCC20,DBLP:conf/ccs/ZhaoZZZZLYYL21,DBLP:conf/ccs/HeX0J23}, which manipulate the malicious application files, causing them to be misclassified as benign by the ML-based AMD systems.

In response to the growing threat of adversarial malware attacks, researchers have proposed various defense methods to enhance the adversarial robustness of ML-based AMD systems~\cite{DBLP:conf/www/LiZYLGC21,DBLP:journals/tdsc/LiCLXXX24,DBLP:journals/tifs/RashidS23,DBLP:journals/tifs/LiL20,DBLP:conf/esorics/GrossePMBM17}.
However, these methods primarily focus on defending against feature space attacks, leaving the more realistic and complex problem space attacks largely unaddressed.
As defined by Pierazzi \textit{et al}.~\cite{DBLP:conf/sp/PierazziPCC20}, the key difference between feature space attacks and problem space attacks lies in the nature of their constraints.
The adversarial example in feature space attacks is constrained to remain close to the original sample in terms of feature similarity.
In contrast, problem space attacks occur in the input domain (e.g., the actual malware files) and must adhere to domain-specific constraints, such as program grammar.

The differing constraints between problem space attacks and feature space attacks make defending against problem space attacks significantly more challenging.
The constraints of adversarial Android malware attacks in the problem space are in the Android application such as the program grammar~\cite{DBLP:conf/ccs/HeX0J23}, whereas in the feature space attacks, the constraints are typically the $\ell_p$ distance~\cite{DBLP:conf/sp/Carlini017}.
This discrepancy makes it challenging to mathematically identify all feasible adversarial regions due to the complexity of maintaining functional consistency, unlike in feature space attacks, where they remain nearby.
Consequently, the generation of adversarial Android malware for problem space defense is typically incomplete.
Additionally, adversarial malware generated in the problem space can exhibit much greater diversity and significant differences in feature vectors compared to the original samples.
In contrast, feature space attacks produce adversarial examples that differ only minimally from the original samples.
This diversity in problem space attacks further complicates the development of robust defenses, as the range of possible adversarial variations is far broader and less predictable.

In terms of the cost, generating the problem space adversarial Android malware is time-consuming~\cite{DBLP:conf/ccs/HeX0J23,DBLP:conf/uss/LucasPLBRS23}.
This process typically involves complex software manipulations, such as decompiling and recompiling APK files, which require significantly more time than feature space attacks, which only perturb numerical feature vectors.
Furthermore, existing state-of-the-art (SOTA) works~\cite{DBLP:conf/www/LiZYLGC21,DBLP:journals/tdsc/LiCLXXX24} adopt more robust model architectures~\cite{DBLP:conf/www/LiZYLGC21} or train more robust model parameters~\cite{DBLP:journals/tdsc/LiCLXXX24} to defend against the adversarial examples by improving the inherent robustness of the ML model.
However, this approach may negatively impact the original malware detection performance of ML-based AMD systems.
This is because strengthening robustness against adversarial perturbations may shift decision boundaries, leading to a less accurate fit for clean data.

To address the aforementioned limitations, we propose a practical \textbf{a}dversarial An\textbf{d}roid malware \textbf{d}efense to enhance the adversarial robustness of ML-based AMD systems, termed \system.
At a high level, \system is designed as a plug-in that revisits samples classified as benign by the original ML-based AMD systems.
This design minimizes the negative impact of \system on the original ML-based AMD systems for two main reasons.
First, the plug-in design employs another detection model to separate the detection ability of adversarial Android malware samples from regular malicious samples.
Consequently, the original malware detection capability of the primary model remains unaffected by the adversarial Android malware samples.
Second, adversarial Android malware attackers are primarily motivated to manipulate malicious samples to be misclassified as benign.
Therefore, \system only revisits samples classified as benign by the underlying model, further reducing the potential influence on the original detection performance of these methods.

To defend against practical adversarial Android malware attacks in the problem space, \system shifts attention from the ML-based AMD systems to the adversarial Android malware attack methods by leveraging the constraints and trade-offs inherent in adversarial Android malware attack methods.
The design of \system is based on the following two key observations.
First, adversarial perturbations in the problem space can only affect a limited set of features within the feature space due to problem space constraints~\cite{DBLP:conf/sp/PierazziPCC20}.
We refer to the features that adversarial perturbations can perturb as the perturbable features and those that cannot be perturbed as the imperturbable features.
Second, adversarial Android malware attack methods target the features that significantly reduce the confidence of the malware label to maximize attack effectiveness and efficiency.
However, the targeted features are likely to introduce semantic incompatibility with the remaining features, which causes an incompatibility between the perturbable features and the imperturbable features in the adversarial malware sample.

Based on these key observations, \system first quantifies the perturbable feature space, which consists of perturbable features, and the imperturbable feature space, which consists of imperturbable features within the feature space of the ML-based AMD system.
The space quantification algorithm is inspired by the Monte Carlo sampling methods~\cite{DBLP:journals/siamcomp/DagumKLR00,DBLP:conf/sigmod/ProcopiucJAM02} by observing the changed features in the space quantification applications.
However, the perturbable and imperturbable feature spaces may not align in dimensions, making it challenging to directly measure incompatibility.
To address this, \system employs encoder models to project the perturbable and imperturbable feature spaces into a common embedding space, allowing the incompatibility score to be computed by measuring the distance between the embedding vectors.
To train the encoder models, \system designs a customized contrastive learning loss utilizing the training dataset of the ML-based AMD system and \textit{pseudo adversarial malware} samples introduced by us.
Finally, any input sample with an incompatibility score exceeding a certain threshold is determined to be malicious.

Upon building the encoder models, the emphasis shifts to determining the threshold for identifying the adversarial Android malware.
To avoid the data snooping pitfall~\cite{DBLP:conf/uss/ArpQPWPWCR22}, \system utilizes a calibration dataset comprising benign and malicious samples.
These samples help measure \system's potential influence on the original detection performance of the ML-based AMD system.
As \system only revisits the benign output of the target model, the potential influence only stems from the true negative samples and false negative samples.
True negative samples may be falsely identified as malicious and false negative samples may be correctly re-identified as malicious by \system.
However, to the best of our knowledge, there are no existing metrics providing the fine-grained measures in two scenarios.
Therefore, \system designs two new metrics termed: True Negative Influence Rate (TNIR) and False Negative Influence Rate (FNIR), respectively.
TNIR measures the proportion of true negative samples incorrectly classified as malicious, while FNIR measures the proportion of false negative samples correctly identified as malicious.
Thus, the threshold calibration algorithm is designed to control TNIR at a specified level while maximizing FNIR.
This approach ensures that the negative impact on the original detection performance is controlled while the positive impact is maximized.

We evaluate \system using a comprehensive Android malware dataset containing 135,859 benign applications and 15,778 malicious applications (151,637 total) dated from 2016 to 2018 and a recent Android malware dataset containing 20,206 malicious applications dated from 2022.
This evaluation demonstrates the defense performance of \system against the SOTA problem space adversarial Android malware attack methods~\cite{DBLP:conf/ccs/HeX0J23,DBLP:journals/compsec/BostaniM24,DBLP:conf/asiaccs/SongLAGKY22} across various ML-based AMD systems.
Additionally, \system significantly improves the false negative performance of ML-based AMD systems while maintaining an acceptable negative influence rate on the true negative samples.
It outperforms the baseline methods (i.e., FD-VAE~\cite{DBLP:conf/www/LiZYLGC21} and adversarial training~\cite{DBLP:journals/tdsc/LiCLXXX24}) in terms of the defense effectiveness against the SOTA adversarial Android malware attacks across various ML-based AMD systems.
In particular, \system achieves a normalized decreased attack success rate of over 95\% against these SOTA adversarial Android malware attack methods in most settings.
Additionally, \system can improve the false negative rate by approximately 50\% while only reducing the true negative rate of the original ML-based AMD system by 1\%.
Furthermore, \system also demonstrates defense effectiveness under two adaptive attack scenarios.
Moreover, \system can be utilized to enhance the adversarial robustness of real-world antivirus solutions (AVs) in VirusTotal.
Specifically, \system achieves over 90\% detection rate to detect the adversarial Android malware samples that reduce the number of detected engines in VirusTotal.
\system still maintains a detection rate of 70\% to 80\% for the adversarial Android malware generated from the more recent malware samples.

The key contributions of this paper are summarized as follows.
We propose \system, which is a practical adversarial Android malware defense framework for enhancing the adversarial robustness of ML-based AMD systems.
In \system, a novel contrastive learning loss is proposed to capture the hierarchical inequality relation of incompatibility scores between pseudo adversarial malware samples, regular malicious samples, and benign samples.
Then, a threshold calibration algorithm that utilizes two new metrics is proposed to optimally select the threshold for incompatibility scores.
To evaluate \system, two new metrics termed TNIR and FNIR are proposed to measure its impact on the original detection performance.

\section{Background}

\subsection{ML-based AMD}

The machine learning method has become a pivotal component in detecting Android malware.
The introduction of learning-based classifiers has significantly improved the effectiveness and generalization capabilities of Android malware detection compared to traditional signature-based methods.
This advancement has led to the widespread deployment of ML-based AMD systems in real-world applications.

To detect the Android malware, ML-based AMD systems typically begin with the utilization of Android application analysis tools, such as Soot~\cite{DBLP:conf/cascon/Vallee-RaiCGHLS99} and Apktool~\cite{Apktool}, to analyze the manifest file and the program code within the Android application.
Based on the analysis results, these methods extract features that are then used by machine learning classifiers to classify.
For instance, Drebin~\cite{DBLP:conf/ndss/ArpSHGR14} extracts syntax features (e.g., permissions) from the manifest file and semantic features (e.g., function calls) from the program code and organizes categorical feature representations.
MaMadroid~\cite{DBLP:conf/ndss/MaricontiOACRS17} only extracts the function call graph features and organizes them as numerical feature representations.
This organized feature vector is then used in the ML model for the classification.
Different ML-based AMD methods may employ different ML classifiers, e.g., Drebin~\cite{DBLP:conf/ndss/ArpSHGR14} employs the SVM and MaMadroid~\cite{DBLP:conf/ndss/MaricontiOACRS17} utilizes the Random Forest (RF) classifier.

We utilize four SOTA and representative ML-based AMD methods in our evaluation.
They are Drebin~\cite{DBLP:conf/ndss/ArpSHGR14}, Drebin-DL~\cite{DBLP:conf/esorics/GrossePMBM17}, MaMadroid~\cite{DBLP:conf/ndss/MaricontiOACRS17}, and APIGraph~\cite{DBLP:conf/ccs/ZhangZZDCZZY20}, according to the previous works~\cite{DBLP:conf/ccs/HeX0J23,DBLP:journals/corr/abs-2402-02953}.
The rationale for choosing these specific models lies in their established performance performance and popularity in research.
For instance, APIGraph achieves an AUROC score of approximately 0.95 in our evaluation (Table~\ref{tab:AMDPerformance}).
Furthermore, these methods have been primary targets in many papers on adversarial malware attacks~\cite{DBLP:conf/ccs/HeX0J23,DBLP:journals/tifs/LiL20,DBLP:conf/uss/0008CWYGYL23}, making them relevant for our analysis.
Moreover, the diversity of their feature spaces and classifiers also ensures a more comprehensive evaluation of our defense mechanisms.

\begin{figure}[t]   
	\centering  
	\includegraphics[width=1\linewidth]{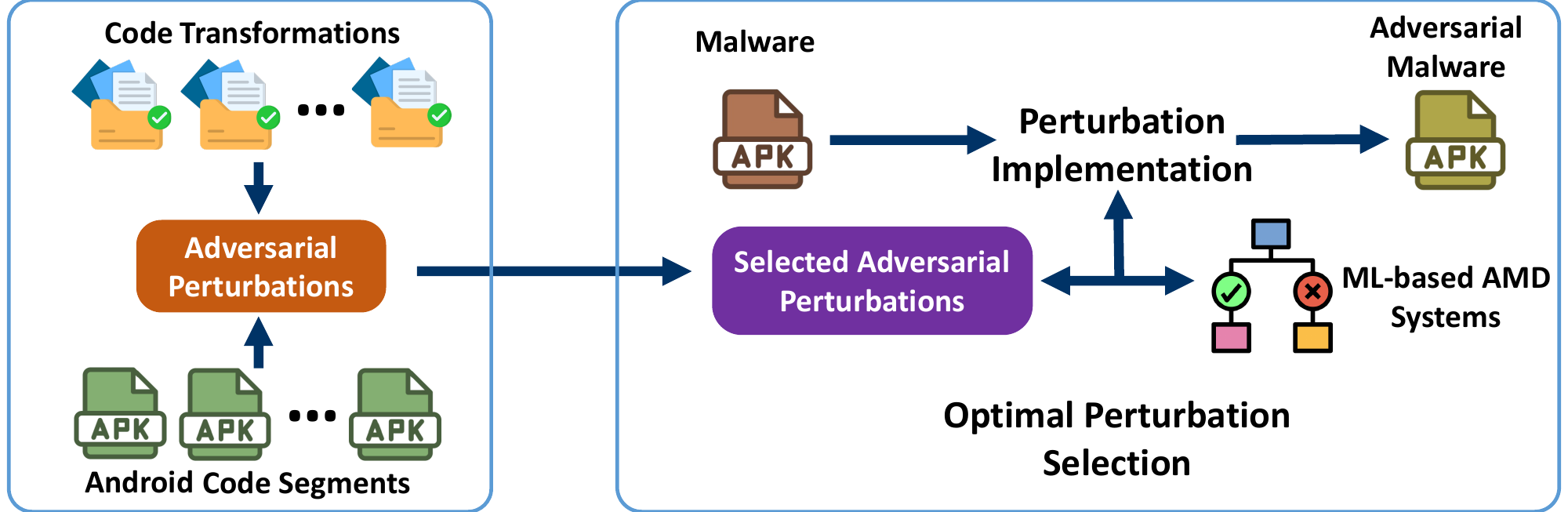}
	\caption{Adversarial Android malware attack framework.}
	\label{fig:aam}
\end{figure}

\subsection{Adversarial Android malware attack}

Adversarial Android malware attacks~\cite{DBLP:conf/esorics/GrossePMBM17,DBLP:conf/sp/PierazziPCC20,DBLP:journals/tifs/0002LWW0N0020,DBLP:conf/ccs/ZhaoZZZZLYYL21,DBLP:conf/uss/0008CWYGYL23,DBLP:conf/ccs/HeX0J23} aim at deliberately manipulating the Android malware samples detected by the ML-based AMD systems to be misclassified as benign samples while preserving the malicious functionality.
Different from adversarial example attacks in image domain~\cite{DBLP:journals/corr/SzegedyZSBEGF13,DBLP:conf/sp/Carlini017,DBLP:conf/sp/ChenJW20}, the adversarial perturbations in Android malware domain must adhere the problem space constraints~\cite{DBLP:conf/sp/PierazziPCC20}.
This necessitates that the adversarial perturbation be implemented in the APK file in accordance with program grammar and maintain functional consistency~\cite{DBLP:conf/ccs/HeX0J23}.

Typically, adversarial Android malware attacks consist of two stages: adversarial perturbation preparation and adversarial perturbation injection, as illustrated in Figure~\ref{fig:aam}.
The adversarial Android malware attack methods first propose the APK manipulation methods that keep the problem space constraints as the potential adversarial perturbation.
These manipulations can take the form of functionality-preserving code transformations~\cite{DBLP:conf/uss/0008CWYGYL23,DBLP:conf/ccs/ZhaoZZZZLYYL21} and the insertion of code segments from Android applications~\cite{DBLP:conf/ccs/HeX0J23,DBLP:conf/sp/PierazziPCC20}.
In the second stage, these manipulation techniques are used during adversarial perturbation injection, where an algorithm is designed to select and inject adversarial perturbations optimally.
The early adversarial Android malware attack methods~\cite{DBLP:conf/esorics/GrossePMBM17,DBLP:conf/sp/PierazziPCC20,DBLP:journals/tifs/0002LWW0N0020,DBLP:conf/ccs/ZhaoZZZZLYYL21,DBLP:conf/uss/0008CWYGYL23,DBLP:conf/sigsoft/SunXTDLWZCN23} assume the attacker has the perfect knowledge or limited knowledge of the ML-based AMD systems, allowing them to leverage model parameters to guide the selection of adversarial perturbations.
For instance, Pierazzi \textit{et al}.~\cite{DBLP:conf/sp/PierazziPCC20} leverage the weights of the SVM classifier to identify the most effective perturbations for injection.
More recent SOTA adversarial malware attack methods consider more practical scenarios where the adversary has zero knowledge of the internal workings of the ML-based AMD systems~\cite{DBLP:conf/ccs/HeX0J23,DBLP:journals/compsec/BostaniM24}.
These methods rely on additional insights from program semantics and query feedback to select optimal adversarial perturbations.
For example, AdvDroidZero~\cite{DBLP:conf/ccs/HeX0J23} employs the perturbation selection tree to choose the most effective adversarial perturbations, demonstrating effectiveness against real-world antivirus engines.

\section{Methodology}


\begin{figure*}[t]   
	\centering  
	\includegraphics[width=0.75\linewidth]{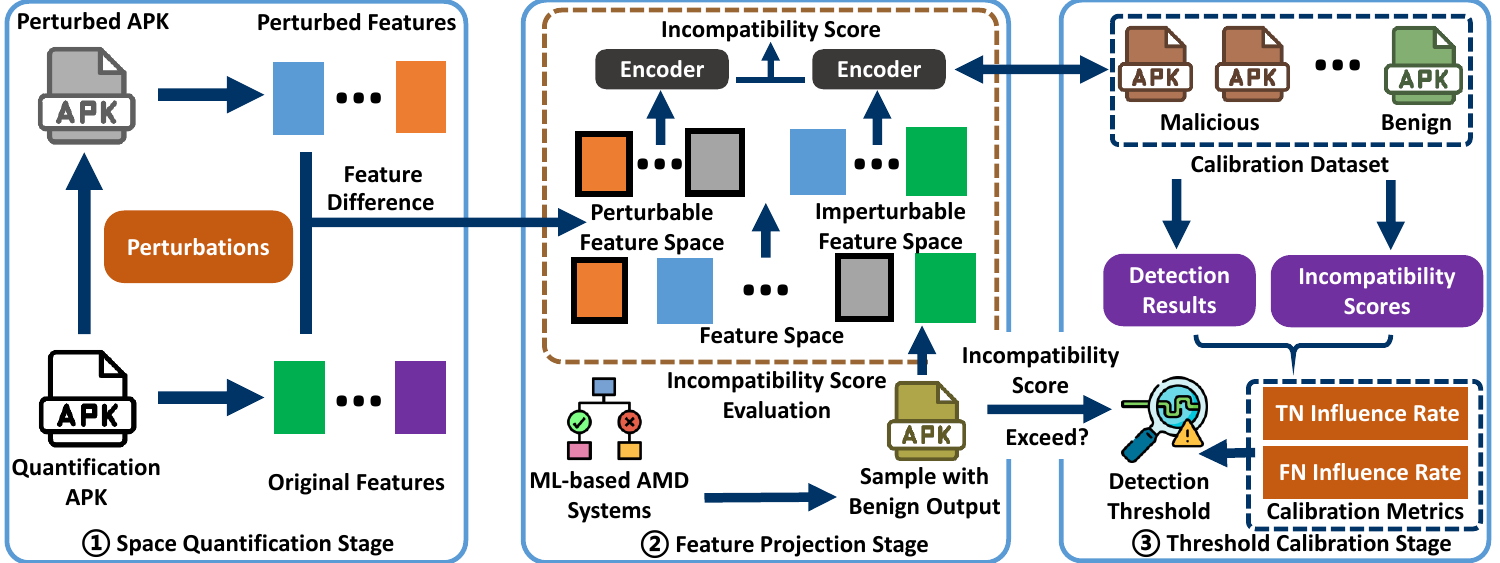}
	\caption{The overview of \system. It operates in three stages: space quantification stage, feature projection stage, and threshold calibration stage.}
	\label{fig:add}
\end{figure*}

\subsection{Threat model}
\label{sec:threatmodel}

We present the threat model of \system by delineating three key components: defender goals, knowledge and capabilities, building upon previous works~\cite{DBLP:conf/satml/ApruzzeseADFPR23,DBLP:journals/corr/abs-1902-06705,DBLP:journals/pr/BiggioR18,DBLP:conf/ccs/HeX0J23}.

\paragraphbe{Defender goals}
The defender aims to detect the adversarial Android malware samples that are misclassified by the ML-based AMD systems.
Given that the adversarial Android malware attackers focus solely on the misclassification of malicious samples as benign samples~\cite{DBLP:journals/corr/abs-1902-06705,DBLP:conf/ccs/HeX0J23}, the defender only needs to detect the adversarial Android malware samples among those classified as benign.
Additionally, the defender seeks to preserve or enhance the original detection performance (e.g., precision, recall) of ML-based AMD systems.
This goal aligns with the need for utility preservation in ML-based AMD systems, emphasizing their detection effectiveness and user experience.

\paragraphbe{Defender knowledge}
We assume that defenders could be the developers of the ML-based AMD systems who are motivated to protect them against the adversarial Android malware attack.
Therefore, the defender has the knowledge of the internals of ML-based AMD systems, including feature spaces, model parameters and the training dataset.
The defender also has the knowledge of the feasible perturbations for the Android malware sample.
This is because the feasible perturbations are determined by the program grammars, which can be found in Android documentation~\cite{AndroidDocumentation} that is available for the defender.
However, in practice, the defender remains uninformed about the specific adversarial Android malware attack methods employed by the adversary.
Additionally, the defender cannot obtain any adversarial Android malware samples generated by the adversary.
This is because the attackers typically do not disclose their adversarial Android malware attack method for selecting the adversarial perturbations.

\paragraphbe{Defender capabilities}
The defender could access the malicious and benign samples identified by the ML-based AMD systems in the APK file format and their corresponding detection results.
This access is feasible because the defender is likely the developer of the ML-based AMD systems.
The ML-based AMD systems process input samples in APK file format and generate corresponding detection results.

\subsection{Key observations}
\label{sec:observations}

The design of our method is inspired by two key observations derived from adversarial Android malware attack methods.

\textbf{Observation \MakeUppercase{\romannumeral 1}}.
Adversarial perturbations in the problem space utilized in adversarial Android malware attack methods must comply with the Android program grammar and maintain functional consistency.
Consequently, the adversarial perturbations could only perturb the feature vector of the malware sample within a restricted space~\cite{DBLP:conf/sp/PierazziPCC20}.
For simplicity, we refer to the portion of feature space of ML-based AMD systems that can be perturbed by the plausible adversarial perturbations in the problem space as the \textbf{perturbable feature space}.
Correspondingly, the features within this space are termed \textbf{perturbable features}.
In contrast, we refer to the portion of feature space of ML-based AMD systems that cannot be perturbed by the plausible adversarial perturbations in the problem space as the \textbf{imperturbable feature space}.
The features within this space are termed \textbf{imperturbable features}.

\textbf{Observation \MakeUppercase{\romannumeral 2}}.
The logic underlying adversarial Android malware attacks is designed to perturb the most vulnerable feature within the feature space of ML-based AMD systems, which most significantly influences the confidence of the malware label.
This is because perturbing the most vulnerable feature can help adversarial Android malware attack to maximize the attack effectiveness and attack efficiency.
However, this approach overlooks the compatibility between the perturbed features and the original features within the malware sample.
This means that the most vulnerable feature is unlikely to align with the most compatible feature.
Consequently, perturbable features are more likely incompatible with imperturbable features within the malware sample.

For example, consider spyware designed for eavesdropping on meetings, disguised as a music application like YouTube Music.
This malware may possess original audio-related features, such as microphone permissions, audio processing functions, and the application name.
However, adversarial Android malware attack methods may inject non-audio-related features, such as GPS or camera permissions, to significantly reduce the confidence of the malware label and evade detection.
These permission features, being perturbable, exhibit semantic incompatibility with the imperturbable features, such as audio processing functions and the application name.


\subsection{Design overview}
\label{sec:overview}

At a high level, \system is an adversarial Android malware defense framework designed as a plug-in for ML-based AMD systems.
It revisits the benign output of the ML-based AMD systems to enhance their robustness against adversarial Android malware attacks.
Building on the key observations outlined in Section~\ref{sec:observations}, \system aims to detect adversarial Android malware samples by evaluating the incompatibility score between their perturbable features and imperturbable features.
It operates through three stages: (1) the space quantification stage, (2) the feature projection stage, and (3) the threshold calibration stage, as shown in Figure~\ref{fig:add}.
In the following, we briefly illustrate every stage, and the algorithmic descriptions for \system are provided in Appendix~\ref{appendix:algorithm}.

\paragraphbe{Space quantification stage}
\system quantifies the perturbable feature space and imperturbable feature space of the ML-based AMD system by leveraging space quantification applications and the feasible perturbation set.
For each perturbation in the feasible perturbation set, \system injects the perturbation into the space quantification application and observes the differences in the feature vector.
The perturbed features are identified as the perturbable features, while the remaining features are categorized as the imperturbable features.

\paragraphbe{Feature projection stage}
Given that the perturbable feature space and imperturbable feature space may differ in dimensions, \system trains two encoder models to project these features.
To train the two encoder models, \system designs the pseudo adversarial Android malware into the customized contrastive learning method for effectiveness and efficiency.
The incompatibility score is calculated based on the distance between the projected perturbable features and imperturbable features.

\paragraphbe{Threshold calibration stage}
\system utilizes a calibration dataset to calibrate the threshold for the incompatibility score and determine the corresponding training epoch for the encoder models.
This dataset consists of benign samples and malicious samples absent from the test set of the target ML-based AMD system.
To determine the threshold, \system introduces two new calibration metrics: True Negative Influence Rate (TNIR) and False Negative Influence Rate (FNIR), which represent the influence on the true negative rate and false negative rate of the target ML-based AMD system in the calibration dataset, respectively.
The threshold is determined by controlling the TNIR at a certain level (e.g., 5\%) while maximizing the FNIR.

        
        
        
        

\subsection{Space quantification stage}

As demonstrated in Section~\ref{sec:observations}, the adversarial perturbations could only affect the perturbable features in the feature space of the ML-based AMD system due to the problem space constraints.
The goal of the space quantification stage is to distinguish between perturbable and imperturbable features.
Although these features can theoretically be identified manually, as they are determined by the ML-based AMD system and Android grammar, the sheer size of the feature space (typically exceeding 10,000 features) makes manual identification impractical.

To address this challenge, \system employs an experimental approach inspired by Monte Carlo sampling~\cite{DBLP:journals/siamcomp/DagumKLR00,DBLP:conf/sigmod/ProcopiucJAM02} to quantify the feature space.
Specifically, \system injects feasible problem space perturbations into Android applications and observes the resulting changes in the feature space.
Features that change are classified as perturbable, while those that remain unaffected are classified as imperturbable.

\system implements this process by applying each perturbation from a predefined malware perturbation set to two space quantification applications.
Observed changes in the application's features are used to identify perturbable features, which are then aggregated to form the perturbable feature space.
Unchanged features constitute the imperturbable feature space.
The malware perturbation set is constructed strictly following the prior adversarial Android malware attack work~\cite{DBLP:conf/ccs/HeX0J23}, including manipulations of both the \textit{AndroidManifest.xml} file and the DEX code.

Regarding space quantification applications, \system incorporates two specific applications: the No Activity application and the Empty Activity application, both generated using the default project templates in Android Studio~\cite{AndroidStudio} without any modifications.
The rationale behind selecting these two applications is as follows.
The problem space perturbations must preserve the functionality of the application.
Thus, they will not influence the original feature of the space quantification application.
This results in that the original features of the space quantification application will never be marked as the perturbable features.
Therefore, an application with fewer features exerts minimal influence on space quantification.
Thus, the No Activity application, devoid of any activity, is chosen for its minimal program semantics.
However, since some problem space malware perturbations may require an activity in the original application to be present, they cannot be injected into the No Activity application, leading to incomplete quantification.
To address this issue, \system also includes the Empty Activity application, which contains only a basic main activity.
The final perturbable feature space is obtained by combining the results from both the No Activity application and the Empty Activity application.
The detailed algorithmic description of the space quantification algorithm can be found in Appendix~\ref{appendix:spacequantify}.


\subsection{Feature projection stage}
\label{sec:featureproject}

Upon acquiring the perturbable feature space and the imperturbable feature space, the emphasis shifts to determining their incompatibility score.
However, these spaces are unlikely to align in dimensions, complicating the calculation of the incompatibility score.
For instance, in our Drebin model~\cite{DBLP:conf/ndss/ArpSHGR14}, the perturbable feature space has 226 dimensions, whereas the imperturbable feature space has 9,774 dimensions.
This stark disparity in feature dimensions renders direct computation of the incompatibility score unsuitable.
To address these challenges, \system employs encoder models designed to project the perturbable feature space and imperturbable feature space into a common lower dimensional space.
The incompatibility score is then calculated based on the distance between the projected features.

\paragraphbe{Model design}
The encoder models are designed to project the perturbable and imperturbable features from high-dimensional spaces to a common low-dimensional space.
This involves two separate encoder models for the perturbable feature space and the imperturbable feature space, respectively.
The architecture of each encoder model is an MLP with the Rectified Linear Unit (ReLU) activation function~\cite{DBLP:conf/nips/KrizhevskySH12} and dropout layer~\cite{DBLP:journals/jmlr/SrivastavaHKSS14}.
The input layer dimensions correspond to the respective dimensions of the perturbable and imperturbable feature spaces.
The output layer dimension is set to a much smaller value, e.g., 32, than the input layer dimension~\cite{DBLP:conf/www/LiZYLGC21,DBLP:journals/nature/LeCunBH15,DBLP:conf/uss/Yang0HCAX021,DBLP:conf/uss/0001D023}.
Considering the large gap between the input and output dimensions, \system adopts an exponential dimension reduction policy to determine the dimensions of the hidden layers.
This policy exponentially reduces the dimensions of the hidden layers, aiming to enhance model performance and optimize memory usage.
More details of the exponential dimension reduction policy can be found in Appendix~\ref{appendix:dimension}.

\paragraphbe{Model training}
The objective of the encoder models is to learn the incompatibility score between the perturbable features and imperturbable features, reflecting it in the distance between the projected features.
Samples with incompatibility scores exceeding the threshold are identified as malicious.
The incompatibility score for a sample $x$ is defined as follows:
\begin{equation}
    \label{equ:score}
    \text{IncalScore}(x) = D(x) = \ell_{2}(\text{EPS}(\text{PS}(x)), \text{EIPS}(\text{IPS}(x))),
\end{equation}
where PS and IPS represent the perturbable feature space and imperturbable feature space, respectively.
EPS and EIPS denote the encoder models for projecting the perturbable feature space and imperturbable feature space, respectively.
$\ell_2$ means the L2-norm loss according to the previous work~\cite{DBLP:conf/sp/ChenJW20}.

As discussed in Section~\ref{sec:overview}, \system revisits only those samples classified as benign by the ML-based AMD system.
There are three scenarios for a sample with a benign output:
\begin{enumerate}

    \item \textbf{True benign samples}: If the sample is indeed benign, \system must retain this classification to avoid increasing false positives, which would degrade the original detection performance of the ML-based AMD system.
    
    \item \textbf{Misclassified malicious samples}: If the sample is actually malicious, \system must identify it as malicious.
    This correction improves the original detection performance by reducing false negatives.

    \item \textbf{Adversarial Android malware samples}: If the sample is an adversarial Android malware, \system must also identify it as malicious due to its inherent malicious functionality.

\end{enumerate}

To achieve this goal, \system employs a customized contrastive learning loss composed of three parts, each addressing one of the corresponding scenarios.
The first part is designed to minimize any negative impact on the original detection performance of the ML-based AMD system.
It ensures that true benign samples are correctly identified to control an increase in the false positive rate at an acceptable level.
The second part aims to enhance the detection capabilities of the ML-based AMD system by ensuring that malicious samples, previously misclassified as benign, are correctly identified.
This improvement reduces the false negative rate.
The third part is focused on effectively detecting adversarial Android malware samples by recognizing their distinct incompatibility features.
These three parts align with the goals of the defender, as discussed in Section~\ref{sec:threatmodel}.

To keep the detection results of the benign sample ($x_b$), the first part of the customized contrastive learning loss minimizes the incompatibility scores of these samples.
This objective is formulated as follows:
\begin{equation}
    D(x_b) = \ell_{2}(\text{EPS}(\text{PS}(x_b)), \text{EIPS}(\text{IPS}(x_b))),
\end{equation}
\begin{equation}
    \label{equ:loss1}
    L_1 = \frac{1}{|\mathbb{D}_b|}\sum_{x_b \in \mathbb{D}_b}D(x_b),
\end{equation}
where PS and IPS, EPS and EIPS retain their meanings as defined in Equation~\ref{equ:score}.
$\mathbb{D}_b$ denotes the set of benign samples in the training set of the ML-based AMD system and $|\mathbb{D}_b|$ represents the cardinality of the set.
Thus, the first part of the customized contrastive learning loss focuses on minimizing the incompatibility scores of all benign samples in the training dataset of the ML-based AMD system.

To improve the original performance of the ML-based AMD system, the second part of the customized contrastive learning loss incorporates a malicious sample ($x_m$) from the training dataset of the ML-based AMD system.
In this case, \system aims to identify these samples them as malicious.
Consequently, this part of the loss function ensures that the incompatibility score of a benign sample is smaller than that of a malicious sample.
This objective can be formulated as:
\begin{equation}
    D(x_m) = \ell_{2}(\text{EPS}(\text{PS}(x_m)), \text{EIPS}(\text{IPS}(x_m))),
\end{equation}
\begin{equation}
    \label{equ:loss2}
    L_2 = \frac{1}{|\mathbb{D}_{bm}|}\sum_{(x_b, x_m) \in \mathbb{D}_{bm}}\max(D(x_b) - D(x_m) + m, 0),
\end{equation}
where PS and IPS, EPS and EIPS keep the same meaning with Equation~\ref{equ:score}.
$\mathbb{D}_{bm}$ is the set that contains the pairs of benign sample and malicious sample in the training set of ML-based AMD system, and $|\mathbb{D}_{bm}|$ represents the cardinality of the set.
$m$ denotes a fixed margin (a hyperparameter).

To achieve effective detection of adversarial Android malware samples, one straightforward solution is to incorporate these samples into the detection model to learn their features. 
The techniques such as adversarial training~\cite{DBLP:journals/corr/GoodfellowSS14,DBLP:conf/nips/ShafahiNG0DSDTG19,DBLP:conf/iclr/ZhuCGSGL20,DBLP:conf/emnlp/YooQ21,DBLP:conf/nips/MaoM0V23} and certified training~\cite{DBLP:conf/iclr/MullerE0V23,DBLP:journals/corr/abs-2306-10426,DBLP:conf/nips/SinghGMPV18,DBLP:conf/nips/ShiWZYH21,DBLP:conf/nips/KrishnanMP20} are widely applied to enhance the adversarial robustness of ML models in image and text domains.
However, in practice, defenders may lack knowledge about the specific adversarial Android malware attack methods employed by adversaries and cannot have access to adversarial Android malware samples generated by them (Section~\ref{sec:threatmodel}).
Additionally, even if defenders know some known adversarial Android malware attack methods, generating adversarial malware in the problem space is computationally expensive~\cite{DBLP:conf/uss/LucasPLBRS23}.
For instance, AdvDroidZero~\cite{DBLP:conf/ccs/HeX0J23} requires approximately 10 minutes to guarantee the success of generating an adversarial Android malware sample.

To address these challenges, \system introduces the concept of \textit{pseudo adversarial malware} to facilitate the training process of the encoder models.
A pseudo adversarial malware sample is a perturbed feature vector of a malware sample that the ML model of the ML-based AMD system misclassifies as benign.
These perturbations occur only in the perturbable feature space due to problem space constraints.
To generate pseudo adversarial malware samples, \system employs a customized random algorithm that randomly perturbs the features of a malicious sample in the perturbable feature space until it is classified as benign.
This random perturbation strategy does not require knowledge of the specific adversarial Android malware attack methods.
Additionally, the pseudo adversarial malware generation algorithm has low time complexity, addressing the computational cost challenge because enumerating all combinations of perturbable features to find all pseudo adversarial malware samples would have exponential complexity.
Although the pseudo adversarial malware generation method may not cover all possible pseudo adversarial malware samples, it can still achieve good effectiveness.
The use of randomness is a well-established approach in other applications and has shown to produce good results, such as random walks in graphs~\cite{DBLP:conf/nips/NikolentzosV20,DBLP:conf/icde/LiYQMJ15,DBLP:conf/kdd/HuangSS21}, Monte Carlo algorithms~\cite{DBLP:journals/siamcomp/DagumKLR00,DBLP:conf/sigmod/ProcopiucJAM02,DBLP:conf/focs/KarpL83}.
The detailed algorithmic description about the pseudo adversarial malware generation algorithm can be found in Appendix~\ref{appendix:pseudo}.

Upon acquiring the pseudo adversarial malware samples, \system incorporates them into the training process of the encoder models to approximate adversarial malware samples.
The goal is to ensure that the incompatibility score of a pseudo adversarial malware sample ($x_{pam}$) is higher than that of a regular malicious sample, as adversarial malware samples exhibit the most severe incompatibility in features.
Consequently, the third part of the customized contrastive learning loss is formulated as follows:
\begin{equation}
    D(x_{pam}) = \ell_{2}(\text{EPS}(\text{PS}(x_{pam})), \text{EIPS}(\text{IPS}(x_{pam}))),
\end{equation}
\begin{equation}
    L_3 = \frac{1}{|\mathbb{D}_{pm}|}\sum_{(x_{pam}, x_m) \in \mathbb{D}_{pm}}\max(D(x_{pam}) - D(x_m) + m, 0),
\end{equation}
where PS and IPS, EPS and EIPS keep the same meaning with Equation~\ref{equ:score}.
$m$ keeps the same meaning with Equation~\ref{equ:loss2}.
$\mathbb{D}_{pm}$ is the set that contains the pairs of pseudo adversarial malware samples and malicious samples, and $|\mathbb{D}_{pm}|$ represents the cardinality of the set.

In summary, the customized contrastive learning loss is the sum of the three parts, and we train the encoder models end-to-end with this loss.
Specifically,
\begin{equation}
    L = \lambda_1L_1 + \lambda_2L_2 + \lambda_3L_3
\end{equation}
where $\lambda_1$, $\lambda_2$, $\lambda_3$ are hyperparameters, a common heuristic in machine learning~\cite{DBLP:conf/uss/0001D023,DBLP:conf/sp/YangCCPTPCW23}.

\subsection{Threshold calibration stage}
\label{sec:calibration}

After building the encoder models for feature projection, \system needs to determine the threshold for the incompatibility score.
If the incompatibility score of an input sample exceeds this threshold, \system identifies it as malicious.
To determine the threshold, \system utilizes a calibration dataset to determine the threshold of the incompatibility score.
As discussed in Section~\ref{sec:threatmodel}, the defender has no knowledge about the specific adversarial Android malware attack methods employed by adversaries and cannot have access to adversarial Android malware samples generated by them.
Therefore, the calibration dataset consists of benign samples and regular malicious samples.
To avoid the data snooping pitfall highlighted by Arp \textit{et al}.~\cite{DBLP:conf/uss/ArpQPWPWCR22}, the calibration dataset needs to be separate from the test set used in the evaluation.

Benign samples correctly classified as benign by the ML-based AMD system may be incorrectly identified as malicious by \system, affecting the true negative rate of the original ML-based AMD system.
Conversely, malicious samples misclassified as benign by the ML-based AMD system may be correctly identified as malicious by \system, affecting the false negative rate.
In response to these cases, \system introduces two new calibration metrics: \textbf{True Negative Influence Rate} (TNIR) and \textbf{False Negative Influence Rate} (FNIR).
TNIR represents the proportion of true negative samples incorrectly identified as malicious by \system (denoted by $N_{itn}$) to the total number of the true negative samples (denoted by $N_{tn}$), i.e., TNIR = $N_{itn} / N_{tn}$.
FNIR represents the proportion of false negative samples correctly identified as malicious by \system (denoted by $N_{cfn}$) to the total number of the false negative samples (denoted by $N_{fn}$), i.e., FNIR = $N_{cfn} / N_{fn}$.
Since a larger TNIR indicates a more negative impact on the original detection performance, \system calibrates the thresholds by controlling TNIR at an acceptable level, e.g., 5\%.
Then \system selects the best threshold that maximizes FNIR, as a larger FNIR signifies a more positive impact on the original detection performance.

\begin{algorithm}[t]
    \footnotesize
    \caption{Threshold Calibration}
    \label{alg:threshold}
    \begin{algorithmic}[1]
        
        \Require{
        Calibration set $\mathbb{D}_c$;
        ML classifier $g$;
        Total training epochs $N$;
        Control rate $K$;
        Perturbable feature encoders \textbf{EPS};
        Imperturbable feature encoders \textbf{EIPS}.
        }
        
        \Ensure{
        Incompatibility score threshold $T$;
        Perturbable feature encoder EPS;
        Imperturbable feature encoder EIPS.
        }
        
        \Statex
        \State $n_b, F, T \leftarrow 0, 0, 0$ \Comment{Initialization.}
        \State $\mathbb{D}_{tn} \leftarrow $ ObtainTN($\mathbb{D}_c$, $g$) \Comment{Obtain the true negative samples.}
        \State $\mathbb{D}_{fn} \leftarrow $ ObtainFN($\mathbb{D}_c$, $g$) \Comment{Obtain the false negative samples.}
        \State $n \leftarrow 0$ \Comment{$n$ represents the training epoch.}
        \While{ $n < N$ }
            \State $\mathbb{S}_{tn} \leftarrow$ ObtainScore($\mathbb{D}_{tn}$, \textbf{EPS}[$n$], \textbf{EIPS}[$n$])   \Comment{Obtain the incompatibility scores of true negative samples.}
            \State $\mathbb{S}_{fn} \leftarrow$ ObtainScore($\mathbb{D}_{fn}$, \textbf{EPS}[$n$], \textbf{EIPS}[$n$])   \Comment{Obtain the incompatibility scores of false negative samples.}
            \State $T_{n} \leftarrow $ Percentile($\mathbb{S}_{tn}$, $K$)    \Comment{Obtain the threshold by controlling the TNIR.}
            \State $F_{n} \leftarrow $ ObtainFNIR($\mathbb{S}_{tn}$, $T_{n}$)    \Comment{Obtain the FNIR.}
            \If {$F_{n} > F$}   \Comment{Update the best epoch.}
                \State $T, F, n_b \leftarrow T_{n}, F_{n}, n$
                \State EPS, EIPS $\leftarrow$ \textbf{EPS}[$n_b$], \textbf{EIPS}[$n_b$]
            \EndIf
            \State $n \leftarrow n + 1$
        \EndWhile
        \State \Return $T$, EPS, EIPS;
        
    \end{algorithmic}
\end{algorithm}

The details of the threshold calibration algorithm can be found in Algorithm~\ref{alg:threshold}.
Initially, the algorithm identifies the true negative samples and false negative samples from the calibration dataset using the ML-based AMD system (lines 2-3).
For each training epoch, it calculates the incompatibility scores for these true negative and false negative samples (lines 6-7).
The algorithm then calibrates the threshold based on the $K$ percentile of the incompatibility scores of the true negative samples to control the TNIR (line 8).
Using this threshold, the algorithm computes the FNIR and selects the epoch with the highest FNIR (lines 9-13).
Ultimately, the corresponding threshold and encoder models for the chosen epoch are determined.

\section{Evaluation}


\subsection{Experimental setup}

\paragraphbe{Implementation details}
\system is a defense framework with three stages designed as a plug-in for enhancing the robustness of ML-based AMD systems against adversarial Android malware attacks.
Our implementation of the prototype of \system is developed using Python, which manages the program logic of the defense process across the three stages. 
For building the encoder models for feature projection, we utilize PyTorch~\cite{DBLP:conf/nips/PaszkeGMLBCKLGA19}.
In our implemented encoder models, we set the dimension of the output layer as 32, a common setting in machine learning~\cite{DBLP:journals/pami/BadrinarayananK17}.
We train the encoder models for a maximum of 50 epochs using the Adam optimizer~\cite{DBLP:journals/corr/KingmaB14} with a learning rate of 0.001.

Regarding the hyperparameters in the customized contrastive learning loss, we set the margin $m$ in both $L_2$ and $L_3$ as 1.0 and set $\lambda_1 = \lambda_2 = \lambda_3 = 1.0$.
To construct the set containing pairs of benign and malicious samples, we uniformly sample a malicious sample from the training dataset of the ML-based AMD system for each benign sample.
Similarly, to build the set containing pairs of pseudo adversarial malware samples and malicious samples, we uniformly sample a pseudo adversarial malware sample for each malicious sample in the training set of the ML-based AMD system.

\paragraphbe{Datasets}
We evaluate the defense effectiveness of \system using two datasets: the primary dataset and the supplementary dataset.
The primary dataset comprises 135,859 and 15,778 malicious samples, totaling 151,637 samples.
This dataset is derived from the previous works~\cite{DBLP:conf/ccs/HeX0J23,DBLP:conf/sp/PierazziPCC20}.
We follow the same procedure as in the previous work~\cite{DBLP:conf/ccs/HeX0J23} for downloading the APKs from AndroZoo~\cite{DBLP:conf/msr/AllixBKT16}.
The samples within the primary dataset are dated between January 2016 and December 2018.
The dataset has already been processed by Pierazzi \textit{et al}. \cite{DBLP:conf/sp/PierazziPCC20}, adhering to the space constraints~\cite{DBLP:conf/uss/PendleburyPJKC19}.

We employ both the time-aware split~\cite{DBLP:conf/uss/PendleburyPJKC19,DBLP:conf/uss/ArpQPWPWCR22} and the random split for the primary dataset.
The time-aware split aims to simulate the practical scenario in which the concept drift~\cite{DBLP:conf/sp/BarberoPPC22} problem naturally exists and should be considered.
Specifically, we utilize samples dated between January 2016 and December 2017 as the training set, samples dated between January 2018 and July 2018 as the calibration set, and samples dated between July 2018 and December 2018 as the test set.
In contrast, the random split aims to remove the effect of goodware/malware evolution, a setting commonly evaluated in previous works~\cite{DBLP:conf/sp/YangCCPTPCW23,DBLP:conf/sp/PierazziPCC20}.
Specifically, we randomly split the primary dataset into the training set, the calibration set, and the test set according to the ratio of about 6:3:1.

To evaluate the temporal effectiveness of \system, we also employ a supplementary dataset downloaded from VirusShare~\cite{virusshare}, consisting of 20,206 malicious samples from 2022.
The rapidly evolving landscape of Android APKs and the fact that the most recent samples in our primary dataset are from 2018 makes this supplementary dataset crucial for assessing the defense effectiveness of \system in a temporal context.
Specifically, we evaluate the defense effectiveness of \system in the context of VirusTotal~\cite{VirusTotal} using the supplementary dataset (Section~\ref{sec:vt}).

\paragraphbe{Target models}
We utilize four SOTA ML-based AMD systems, namely Drebin~\cite{DBLP:conf/ndss/ArpSHGR14}, Drebin-DL~\cite{DBLP:conf/esorics/GrossePMBM17}, MaMadroid~\cite{DBLP:conf/ndss/MaricontiOACRS17}, and APIGraph~\cite{DBLP:conf/ccs/ZhangZZDCZZY20}, as our target ML-based AMD systems.
To ensure the fidelity of our implementations, we strictly follow the descriptions and configurations provided in the previous works, as well as the corresponding open-source code~\cite{DBLP:conf/ccs/HeX0J23,DBLP:conf/sp/PierazziPCC20,DBLP:conf/sp/YangCCPTPCW23}.
In line with previous studies, we reduce the feature space to improve both training efficiency and model robustness for Drebin and Drebin-DL, as suggested by Demontis \textit{et al}.~\cite{DBLP:journals/tdsc/DemontisMBMARCG19}.
Specifically, we use the Linear SVM $\ell_2$ regularizer to select the top 10,000 features following the previous work~\cite{DBLP:conf/sp/YangCCPTPCW23}, which also maintains a similar accuracy as using the full feature set.
Regarding the ML classifier, we employ SVM with the linear kernel for Drebin and APIGraph, while for Drebin-DL, we utilize a two-layer MLP as the target classifier.
For MaMadroid, we use RF as the target ML classifier.
These target models encompass diverse feature types and machine learning classifiers, broadly representing the SOTA ML-based AMD systems.
Further details regarding the implementation and the detection performance of these target models can be found in Appendix~\ref{appendix:targetmodel}.

\paragraphbe{Attack methods}
We select three SOTA adversarial Android malware attack methods, namely AdvDroidZero~\cite{DBLP:conf/ccs/HeX0J23}, MAB~\cite{DBLP:conf/asiaccs/SongLAGKY22}, and RA~\cite{DBLP:journals/compsec/BostaniM24}, as our target adversarial Android malware attack methods.
To ensure implementation fidelity, we follow the descriptions and open-source code in their respective publications.
The rationale behind selecting the three methods can be attributed to their practicality and effectiveness against the ML-based AMD systems.
For instance, AdvDroidZero achieves approximately 90\% ASR against the real-world antivirus solutions~\cite{DBLP:conf/ccs/HeX0J23}.
To generate the adversarial Android malware samples on the target models for evaluation, we randomly select 1,000 true positive malware samples from the test set in the primary dataset of the corresponding ML-based AMD systems, meaning these samples are accurately classified as malicious.
We then apply the three adversarial Android malware attack methods, each with a query budget of 10, against the target models.
We report the ASR and the actual number of successfully generated adversarial Android malware in each case, detailed in Appendix~\ref{appendix:attackmethod}.

\begin{table*}[t]\centering
	\caption{The defense performance of \system measured by the NDASR. TNIR@K represents controlling the TNIR at K\% in the calibration dataset. ADZ represents AdvDroidZero.}
	\begin{tabular}[centering,width=0.5*\linewidth]{@{}C{1.2cm}C{1.6cm}C{1.3cm}C{1.0cm}C{1.0cm}C{1.0cm}C{1.0cm}C{1.0cm}C{1.0cm}C{1.0cm}C{1.0cm}C{1.0cm}@{}}
		\toprule
            \multirow{2}{*}{\begin{tabular}[c]{@{}c@{}}	Dataset \\ Split \end{tabular}} & \multirow{2}{*}{\begin{tabular}[c]{@{}c@{}}	AMD \\ Method \end{tabular}} & \multirow{2}{*}{\begin{tabular}[c]{@{}c@{}}	Classifier \end{tabular}} & \multicolumn{3}{c}{TNIR@10} & \multicolumn{3}{c}{TNIR@5} & \multicolumn{3}{c}{TNIR@1} \\ \cmidrule(l){4-6} \cmidrule(l){7-9} \cmidrule(l){10-12}
             & & & ADZ & MAB & RA & ADZ & MAB & RA & ADZ & MAB & RA \\
		\midrule
         \multirow{5}{*}{Time} & Drebin & SVM & 97.53\% & 95.10\% & 96.94\% & 94.32\% & 88.66\% & 91.53\% & 80.49\% & 72.42\% & 75.06\% \\
         \cmidrule(l){4-6} \cmidrule(l){7-9} \cmidrule(l){10-12}
         & Drebin-DL & MLP & 98.55\% & 97.75\% & 97.15\% & 94.77\% & 91.55\% & 94.60\% & 63.66\% & 47.89\% & 46.69\% \\
         \cmidrule(l){4-6} \cmidrule(l){7-9} \cmidrule(l){10-12}
         & APIGraph & SVM & 96.41\% & 97.47\% & 98.13\% & 92.40\% & 89.90\% & 92.31\% & 59.70\% & 57.83\% & 57.81\% \\
         \cmidrule(l){4-6} \cmidrule(l){7-9} \cmidrule(l){10-12}
         & MaMadroid & RF & 98.40\% & 98.02\% & 98.10\% & 94.79\% & 95.12\% & 95.65\% & 83.82\% & 85.64\% & 84.78\% \\ 
        \midrule
        \multirow{5}{*}{Random} & Drebin & SVM & 99.20\% & 99.19\% & 99.24\% & 96.65\% & 96.96\% & 96.23\% & 86.74\% & 85.85\% & 85.66\% \\
        \cmidrule(l){4-6} \cmidrule(l){7-9} \cmidrule(l){10-12}
         & Drebin-DL & MLP & 99.24\% & 99.58\% & 99.38\% & 97.52\% & 98.34\% & 98.15\% & 89.52\% & 89.63\% & 89.89\% \\
         \cmidrule(l){4-6} \cmidrule(l){7-9} \cmidrule(l){10-12}
         & APIGraph & SVM & 98.71\% & 97.85\% & 99.30\% & 96.94\% & 96.68\% & 97.02\% & 82.12\% & 81.54\% & 82.84\% \\
         \cmidrule(l){4-6} \cmidrule(l){7-9} \cmidrule(l){10-12}
         & MaMadroid & RF & 99.21\% & 99.29\% & 99.54\% & 98.32\% & 98.12\% & 98.15\% & 93.38\% & 92.71\% & 91.67\% \\
        \bottomrule
  	\end{tabular}
 
	\label{tab:defense}
\end{table*}

\paragraphbe{Metrics}
As discussed in Section~\ref{sec:threatmodel}, the defender aims to detect the adversarial Android malware and preserve or even improve the original detection performance of ML-based AMD systems.
To measure the first goal, we employ the metric termed as \textbf{normalized decreased attack success rate} (NDASR).
NDASR represents the normalized proportion of the decreased attack success rate (denoted by $ASR_d$) to the original attack success rate (denoted by $ASR_o$), i.e., NDASR = $ASR_d/ASR_o$.
This normalization is necessary because different target models and attack methods exhibit varying attack success rates, making it inappropriate to directly compare the decreased attack success rate across different settings.
To measure the second goal, we propose two new metrics: TNIR and FNIR, which are introduced in Section~\ref{sec:calibration}.
The underlying reasons for the two new metrics are illustrated as follows.
\system only revisits the benign output (negative label) of the ML-based AMD system.
This means that it may influence either true negative samples or false negative samples.
Therefore, the TNIR and FNIR provide more fine-grained measures of \system than the common metrics, e.g., F1, as common metrics also influenced by the malicious output (positive label) of the ML-based AMD system.
More specifically, TNIR indicates the negative influence of \system on the original ML-based AMD system, while FNIR reflects the positive influence of \system on the original ML-based AMD system.

\paragraphbe{Baseline methods}
We implement and compare two baseline methods with \system to evaluate their defense effectiveness against the target adversarial Android malware attack methods.
The two baseline methods are FD-VAE~\cite{DBLP:conf/www/LiZYLGC21} and adversarial training~\cite{DBLP:journals/tdsc/LiCLXXX24}.
Both baseline methods are designed to enhance model robustness rather than perform input-level detection. 
Therefore, we adapt them to our evaluation setting for a fair comparison.
For FD-VAE, we use the training dataset of the target ML-based AMD systems to train the model.
To ensure implementation fidelity, we strictly adhere to the descriptions and configurations provided in the original paper and corresponding open-source code.
Once the FD-VAE model is trained, we use it to detect adversarial Android malware samples generated against the target ML-based AMD systems.
Regarding adversarial training, we generate adversarial examples for training using a pseudo-adversarial malware generation method on the training dataset of the target ML-based AMD systems.
The rationale for this approach is twofold.
First, the use of pseudo-adversarial malware generation ensures a fair comparison between \system and adversarial training.
Second, pseudo-adversarial generation methods efficiently create adversarial examples in the feature space across different ML classifiers.
In contrast, gradient-based methods such as Projected Gradient Descent~\cite{DBLP:conf/nips/PintorDSDCBR22,DBLP:conf/sp/Carlini017} are only applicable to neural networks, such as MLPs.
Moreover, in practical scenarios, defenders typically only have access to the training dataset of the target ML-based AMD systems.
Therefore, we use the training dataset to generate adversarial examples for adversarial training, ensuring alignment with real-world constraints.

\paragraphbe{Experimental environments}
We run all experiments on a Ubuntu 20.04 server with 251G memory and 39G swap memory, 2 Intel(R) Xeon(R) Gold 6346 CPUs and 8 NVIDIA RTX 3090 GPUs.

\subsection{Defense performance}
\label{sec:defper}

To assess the performance of \system, we apply it to defend the aforementioned target attack methods on the target models.
Initially, we evaluate the defense effectiveness of \system compared to the two baseline methods.
Subsequently, we assess the effectiveness of the threshold calibration stage by comparing it with alternative thresholds.

\paragraphbe{Defense effectiveness}
To evaluate the defense effectiveness, we first generate the adversarial Android malware samples using the target attack methods on the target models, then apply \system to detect these samples.
This setting is based on the fact that the defender needs to detect the adversarial Android malware samples generated before the defense method is applied.
To be specific, we randomly select 1,000 true positive malicious samples from the test set in the primary dataset for each target model and use the target attack methods to generate adversarial Android malware samples.
The generation results are detailed in Appendix~\ref{appendix:attackmethod}.
Subsequently, we build \system to detect the generated adversarial Android malware in each target model and measure the evaluation metrics.
Because of the space quantification stage, \system cannot directly be applied to the MaMadroid model with the RF classifier.
Therefore, we utilize the transferability strategy of \system to defend the MaMadroid model with the RF classifier.
We build \system in the target model with Drebin feature space and SVM classifier and detect the adversarial Android malware generated against the MaMadroid model with the RF classifier.

Table~\ref{tab:defense} presents the defense effectiveness of \system measured by the NDASR against SOTA adversarial Android malware attack methods in the target models under varying controlled TNIR.
\system achieves an NDASR of about 95\% against all adversarial Android malware attack methods, regardless of the target models when the TNIR is controlled at 5\%.
Even in the setting of the MaMadroid model with the RF classifier, \system still achieves an NDASR of about 95\% in most settings.
This means that \system has strong defense effectiveness in the transferability setting.
This demonstrates that \system exhibits strong defense effectiveness against the adversarial Android malware attack regardless of the setting.

Table~\ref{tab:influence} presents the influence on the original detection performance of \system measured by the TNIR and FNIR in the target models under varying controlled TNIR.
\system improves the FNIR of the original ML-based AMD system at about 70\% when the TNIR is controlled at 5\% and the FNIR of about 50\% when the TNIR is controlled at 1\%.
This observation demonstrates the effectiveness of \system in correcting the false negative samples of the original ML-based AMD systems.

The performance of \system increases as the controlled TNIR in the calibration set increases, as shown in Table~\ref{tab:defense}.
For example, with the Drebin model with the SVM classifier trained on the time-aware split dataset, \system achieves the NDASR at 80.49\% when the TNIR is controlled at 1\%, while achieving the NDASR at 97.53\% when the TNIR is controlled at 10\%.
However, increasing the controlled TNIR also potentially increases the negative influence on the original performance of ML-based AMD systems.
In practice, \system can employ any controlled TNIR depending on the practical demand of the defender.

Another observation is that target models trained on the randomly split dataset demonstrate more substantial defense effectiveness with \system than those trained on the time-aware split dataset.
For instance, with the Drebin-DL model with the MLP classifier under TNIR@1 against AdvDroidZero, \system achieves the NDASR at 89.52\% for the model trained on the random split dataset, while the NDASR is 63.66\% for the model trained on the time-aware split dataset.
This can be attributed to the impact of concept drift in malware data.
Since the encoder models are also trained on the training dataset of the ML-based AMD systems, the distribution shift also influences the detection effectiveness of \system.


\begin{table}[t]\centering
	\caption{The influence on original detection performance of \system measured by the FNIR and TNIR. TNIR@K represents controlling the TNIR at K\% in the calibration dataset.}
	\begin{tabular}[centering,width=0.5*\linewidth]{@{}C{1.2cm}C{1.6cm}C{0.8cm}C{1.0cm}C{0.8cm}C{1.0cm}@{}}
		\toprule
            \multirow{2}{*}{\begin{tabular}[c]{@{}c@{}}	Dataset \\ Split \end{tabular}} & \multirow{2}{*}{\begin{tabular}[c]{@{}c@{}}	AMD \\ Method \end{tabular}} & \multicolumn{2}{c}{TNIR@5} & \multicolumn{2}{c}{TNIR@1}  \\ \cmidrule(l){3-4} \cmidrule(l){5-6}
             & & TNIR & FNIR & TNIR & FNIR \\
		\midrule
         \multirow{5}{*}{Time} & Drebin & 5.09\% & 57.36\% & 1.01\% & 53.96\% \\
         \cmidrule(l){3-4} \cmidrule(l){5-6}
         & Drebin-DL & 4.89\% & 47.22\% & 0.38\% & 30.02\% \\
         \cmidrule(l){3-4} \cmidrule(l){5-6}
         & APIGraph & 4.66\% & 56.14\% & 0.29\% & 41.31\% \\
         \cmidrule(l){3-4} \cmidrule(l){5-6}
         & MaMadroid & 5.37\% & 39.84\% & 1.84\% & 34.38\% \\ 
        \midrule
        \multirow{5}{*}{Random} & Drebin & 4.93\% & 75.54\% & 1.06\% & 50.54\% \\
        \cmidrule(l){3-4} \cmidrule(l){5-6}
         & Drebin-DL & 4.98\% & 73.55\% & 0.91\% & 51.61\% \\
        \cmidrule(l){3-4} \cmidrule(l){5-6}
         & APIGraph & 5.02\% & 72.54\% & 1.01\% & 49.74\% \\
        \cmidrule(l){3-4} \cmidrule(l){5-6}
         & MaMadroid & 5.50\% & 82.11\% & 1.40\% & 63.01\% \\
        \bottomrule
  	\end{tabular}
 
	\label{tab:influence}
\end{table}

\begin{table}[t]\centering
	\caption{The defense performance of the two baseline methods against AdvDroidZero across different target models measured by the NDASR. The NDASR of \system is measured under the TNIR@5.}
	\begin{tabular}[centering,width=0.5*\linewidth]{@{}C{1.3cm}C{1.6cm}C{1.0cm}C{1.5cm}C{1.5cm}@{}}
		\toprule
            Dataset Split & AMD Method & \system & FD-VAE & Adversarial Training \\
		\midrule
         \multirow{5}{*}{Time} & Drebin & \textbf{94.32\%} & 74.17\% & 26.25\% \\
         \cmidrule(l){3-5}
         & Drebin-DL & \textbf{94.77\%} & 70.35\% & 46.51\% \\
         \cmidrule(l){3-5}
         & APIGraph & \textbf{92.40\%} & 77.00\% & 21.10\% \\
         \cmidrule(l){3-5}
         & MaMadroid & \textbf{94.79\%} & 81.82\% & 1.60\% \\ 
        \midrule
        \multirow{5}{*}{Random} & Drebin & \textbf{96.65\%} & 61.02\% & 17.25\% \\
         \cmidrule(l){3-5}
         & Drebin-DL & \textbf{97.52\%} & 52.95\% & 25.52\% \\
         \cmidrule(l){3-5}
         & APIGraph & \textbf{96.94\%} & 59.26\% & 25.52\% \\
         \cmidrule(l){3-5}
         & MaMadroid & \textbf{98.32\%} & 75.31\% & 5.05\% \\ 
        \bottomrule
  	\end{tabular}
 
	\label{tab:baseline}
\end{table}

\paragraphbe{Baseline methods}
To compare with \system, we evaluate the defense effectiveness of the baseline methods.
Specifically, we evaluate their effectiveness against the AdvDroidZero~\cite{DBLP:conf/ccs/HeX0J23}, which is the SOTA adversarial Android malware attack method under the 10 query budgets in the target models.
Table~\ref{tab:baseline} presents the defense performance of the baseline methods compared with \system.

\system consistently outperforms both baseline methods across all settings, including different dataset split methods, feature spaces, and ML classifiers.
The FD-VAE~\cite{DBLP:conf/www/LiZYLGC21} model achieves an NDASR of approximately 70\% in most settings, which indicates limited effectiveness against adversarial Android malware attacks, particularly when the target model is trained on a randomly split dataset.
Adversarial training proves even less effective, especially when the target model uses the MaMadroid feature set and a Random Forest (RF) classifier, where it achieves an NDASR of only 5\%.
A potential explanation for this poor performance is the constraints imposed by problem-space attacks, making it more difficult for adversarially trained models to classify the generated adversarial malware samples correctly.



\paragraphbe{Threshold calibration effectiveness}
As the threshold for determining the incompatibility score can be viewed as a hyperparameter in practice, we evaluate the effectiveness of our threshold calibration algorithm.
As discussed in Section~\ref{sec:calibration}, different training epochs of the encoder models yield different thresholds under a given controlled TNIR.
Thus, we evaluate our calibrated threshold against the average results of all possible thresholds.
Specifically, we report the average defense performance against AdvDroidZero across all possible thresholds, measured by the three evaluation metrics.
The detailed results of the average defense performance of all thresholds can be found in Appendix~\ref{appendix:avg_defense}.
This result shows that the threshold determined by the threshold calibration stage outperforms the average detection performance of all thresholds, underscoring the effectiveness of the threshold calibration stage.


\subsection{Possible adaptive attacks}
According to the Kerckhoffs' principle~\cite{DBLP:reference/crypt/Petitcolas11}, a defense method must remain effective even under adaptive attacks.
Therefore, we evaluate the defense effectiveness of \system under two adaptive attack scenarios.

\paragraphbe{Adaptive attack \MakeUppercase{\romannumeral 1}}
In practice, the final detection results of input samples are determined by both the original ML-based AMD system and \system.
Therefore, the original ML-based AMD system and \system would be viewed as a whole pipeline by the attackers for obtaining the combined detection outputs as feedback.
This is also a realistic setting since the attackers usually have zero knowledge about the whole pipeline~\cite{DBLP:conf/iclr/BrendelRB18}.
Consequently, we design this adaptive attack method following the previous work~\cite{DBLP:conf/uss/LiDJZLY020} under this setting.
Specifically, we randomly sample 100 true positive malware samples from the primary dataset, ensuring they are classified as malicious by the entire pipeline.
We then employ AdvDroidZero with a query budget of 10 to attack these samples, using the combined outputs as feedback.

\paragraphbe{Adaptive attack \MakeUppercase{\romannumeral 2}}
In this scenario, we consider a more powerful adaptive attacker who can access the original ML-based AMD systems and \system separately.
This adaptive attack method first uses adversarial Android attack methods to generate multiple variants of adversarial malware samples, exploiting the inherent randomness in the attack method.
It then computes the incompatibility scores for these generated variants and selects the sample with the lowest score.
The attack is successful if the selected sample's incompatibility score is below the threshold.
Specifically, we randomly select 100 true positive malware samples from the primary dataset, ensuring they are classified as malicious by \system.
We then employ AdvDroidZero with a query budget of 10 to generate 10 different variants of each malicious sample and select the one with the lowest incompatibility score.

\begin{table}[t]\centering
	\caption{The attack performance of the adaptive attack methods against \system by the TNIR controlled at 5\% measured by ASR.}
 
	\begin{tabular}[centering,width=0.5*\linewidth]{@{}C{1.3cm}C{1.6cm}C{1.0cm}C{1.5cm}C{1.5cm}@{}}
		\toprule
            Dataset Split & AMD Method & Classifier & Adaptive Attack \MakeUppercase{\romannumeral 1} & Adaptive Attack \MakeUppercase{\romannumeral 2} \\
		\midrule
        \multirow{4}{*}{Time} & Drebin & SVM & 1.37\% & 0.00\% \\
         & Drebin-DL & MLP & 4.40\% & 3.37\%  \\
         & APIGraph & SVM & 6.00\% & 15.00\% \\
         & MaMadroid & RF & 11.48\% & 12.00\% \\
         \midrule
        \multirow{4}{*}{Random} & Drebin & SVM & 2.00\% & 2.00\% \\
         & Drebin-DL & MLP & 2.20\% & 1.06\% \\
         & APIGraph & SVM & 0.00\% & 4.00\% \\
         & MaMadroid & RF & 2.56\% & 3.00\% \\
        \bottomrule
  	\end{tabular}
 
	\label{tab:adaptive}
\end{table}

Table~\ref{tab:adaptive} presents the attack effectiveness of these adaptive attack methods against \system under TNIR@5.
These adaptive attacks are primarily ineffective against \system, with attack success rates typically below 10\%.
This can be explained by the fact that \system utilizes the inherent trade-off to defend against adversarial Android malware attacks (Section~\ref{sec:observations}).
Adaptive attack methods still face trade-offs in generating adversarial malware samples, failing to achieve significant attack effectiveness.
This demonstrates that \system has strong defense capability against the adaptive attacks.

\subsection{Real-world AVs}
\label{sec:vt}

To evaluate the defense performance of \system in the context of real-world AVs, we evaluate it on VirusTotal, which hosts over 70 commercial AVs.
To generate the adversarial Android malware samples on the real-world AVs, we follow the procedure in AdvDroidZero~\cite{DBLP:conf/ccs/HeX0J23} by uploading the applications to VirusTotal through the VirusTotal API~\cite{VirusTotalAPI}.
Specifically, we randomly sample 1,000 malicious samples in the test sets of the primary dataset and set the 10 query budget for each malicious sample for generation.
As a result, we successfully generate 805 adversarial Android malware samples which reduces the number of detected antivirus engines in the test set obtained by the time-aware split and 819 adversarial Android malware samples in the test set obtained by the random split.

As discussed in Section~\ref{sec:threatmodel}, the direct utilization of \system needs to have the perfect knowledge of the ML-based AMD system.
However, the internals of the real-world AVs remain undisclosed to us.
To bridge this gap, we leverage the transferability of \system by utilizing the trained encoder models and calibrated threshold within our target models.
Specifically, we generate the pseudo adversarial malware samples from our local model and then utilize them to train the encoder.
We then apply the trained encoder to detect the adversarial examples generated against the real-world AVs.
Subsequently, we compute the NDASR in every case, as shown in Table~\ref{tab:vtres}.
\system achieves an NDASR over 85\% in defending adversarial Android malware samples in most cases.
Additionally, \system achieves an NDASR of nearly 100\% regardless of the target models when the TNIR is controlled at 10\%.
This demonstrates that \system exhibits strong defense effectiveness against adversarial Android malware in the context of real-world antivirus solutions, indicating its practicality.

\begin{table}[t]\centering
	\caption{The NDASR of \system across different target models and TNIRs within the context of real-world antivirus solutions in the primary dataset. TNIR@K means controlling the TNIR at K\% in the calibration dataset.}
 
	\begin{tabular}[centering,width=0.5*\linewidth]{@{}C{1.3cm}C{1.6cm}C{1.3cm}C{1.3cm}C{1.3cm}@{}}
		\toprule
            Dataset Split & AMD Method & TNIR@10 & TNIR@5 & TNIR@1 \\
		\midrule
        \multirow{3}{*}{Time} & Drebin & 98.38\% & 96.15\% & 85.22\% \\
         & Drebin-DL & 97.89\% & 96.27\% & 75.53\% \\
         & APIGraph & 98.88\% & 96.15\% & 77.76\% \\
         \midrule
        \multirow{3}{*}{Random} & Drebin & 98.38\% & 96.15\% & 85.22\% \\
         & Drebin-DL & 99.63\% & 98.90\% & 95.97\% \\
         & APIGraph & 99.14\% & 98.29\% & 88.77\% \\
        \bottomrule
  	\end{tabular}
 
	\label{tab:vtres}
\end{table}

\begin{table}[t]\centering
	\caption{The NDASR of \system across different target models and TNIRs within the context of real-world antivirus solutions in the supplementary dataset. TNIR@K means controlling the TNIR at K\% in the calibration dataset.}
 
	\begin{tabular}[centering,width=0.5*\linewidth]{@{}C{1.3cm}C{1.6cm}C{1.3cm}C{1.3cm}C{1.3cm}@{}}
		\toprule
            Dataset Split & AMD Method & TNIR@10 & TNIR@5 & TNIR@1 \\
		\midrule
        \multirow{3}{*}{Time} & Drebin & 86.28\% & 75.65\% & 59.94\% \\
         & Drebin-DL & 85.98\% & 77.19\% & 50.69\% \\
         & APIGraph & 87.52\% & 78.43\% & 55.47\% \\
         \midrule
        \multirow{3}{*}{Random} & Drebin & 84.59\% & 79.35\% & 52.54\% \\
         & Drebin-DL & 83.36\% & 78.81\% & 60.40\% \\
         & APIGraph & 86.44\% & 76.12\% & 43.76\% \\
        \bottomrule
  	\end{tabular}
 
	\label{tab:vttimeres}
\end{table}

\paragraphbe{Temporal Effectiveness}
To assess the effectiveness of \system over time, we evaluate it on VirusTotal using the supplementary dataset.
Specifically, We randomly sample 1,000 malware samples from the supplementary dataset and employ the same strategy that is used for the malware samples in the primary dataset to generate the adversarial Android malware samples.
In total, we successfully generate 649 adversarial Android malware samples that reduce the number of detected antivirus engines.
We then adopt the same procedure in the primary dataset to detect these generated adversarial Android malware samples.
As shown in Table~\ref{tab:vttimeres}, \system maintained an NDASR between 70\% and 90\% for the adversarial Android malware samples in most cases within the supplementary dataset, underlining its effectiveness over time.

\section{Discussion}


In this study, we introduce an adversarial Android malware defense framework, \system, designed as a plug-in for enhancing the robustness of ML-based AMD systems against adversarial malware attacks.
Our work aims to offer a practical solution and valuable insights into the field of AMD for defending potential adversarial malware attacks.
In the following, we discuss some limitations of our approach and outline potential future research directions. 

\paragraphbe{Extension to other features}
The design of \system is based on the key observation that the adversarial perturbations primarily affect the perturbable feature space.
Thus, the perturbable features in the feature space of the ML-based AMD system need to be quantified.
This will require the ML-based AMD systems to utilize the categorical encoding approaches, which are the most frequently feature encoding methods in ML-based AMD systems according to a recent survey~\cite{DBLP:journals/csur/LiuTLL23}.
However, in some cases~\cite{DBLP:conf/ndss/MaricontiOACRS17}, the perturbable feature space cannot be easily quantified, which can limit the effectiveness of \system.
Additionally, \system assumes that the defender has perfect knowledge of the ML-based AMD system, which can also limit its adaptability.
To adapt \system to these scenarios, the transferability strategy can be implemented (MaMadroid model results in Section~\ref{sec:defper} and real-world AVs results in Section~\ref{sec:vt}).
This technique can provide a practical solution for adapting \system to other ML-based AMD systems and demonstrate its strong transferability effectiveness.
Nonetheless, we acknowledge that the adaptability of \system is limited due to the constraints in the feature type.
In future research, we plan to tackle these specific cases.

\paragraphbe{Influence from concept drift}
The \system framework relies on the training set of the ML-based AMD system.
However, the rapid evolution of malware samples introduces the concept drift problem, which can affect the defense effectiveness of \system.
Concept drift introduces the distribution shift of the statistical properties of the malware samples over time.
This issue has been studied extensively over the years~\cite{DBLP:conf/sp/BarberoPPC22}, with many orthogonal solutions to \system.
For instance, the encoder models of \system can be periodically retrained to adapt to new data.
We plan to explore the impact of concept drift on \system in future research.




\section{Related work}

Adversarial examples have recently garnered significant attention in the malware domain.
Numerous attack algorithms~\cite{DBLP:conf/sp/PierazziPCC20,DBLP:journals/tifs/LiL20,DBLP:conf/ccs/ZhaoZZZZLYYL21,DBLP:conf/ccs/HeX0J23,DBLP:conf/asiaccs/LucasSBRS21} have been developed to generate the adversarial malware samples to evade the detection.
Correspondingly, the defense approaches against such attacks are emerging~\cite{DBLP:conf/uss/LucasPLBRS23,DBLP:conf/www/LiZYLGC21,DBLP:journals/tdsc/LiCLXXX24,DBLP:journals/corr/abs-2312-06423,DBLP:conf/infocom/LingWDQZZMWWJ22,DBLP:journals/tifs/RashidS23,DBLP:conf/uss/Chen0SJ20}.
For instance, Chen \textit{et al}.~\cite{DBLP:conf/uss/Chen0SJ20} propose distance metrics in the PDF tree structure and verifiable robust training methods for training PDF malware classifiers.
Lucas \textit{et al}.~\cite{DBLP:conf/uss/LucasPLBRS23} improve the efficiency and scalability of creating adversarial Windows PE malware to make adversarial training practical using distributed implementations.

In the Android malware domain, Li \textit{et al}.~\cite{DBLP:conf/www/LiZYLGC21} employ a Variational Autoencoder (VAE) to disentangle features of different classes, thereby enhancing the detection robustness.
Li \textit{et al}.~\cite{DBLP:journals/tifs/LiL20} explore the robustness of utilizing the ensemble learning algorithms in defending adversarial Android malware.
Recently, Rashid \textit{et al}.~\cite{DBLP:journals/tifs/RashidS23} present a stateful defense that utilizes indicators within the query history to detect the query-based attacks at the user level.
Zhou \textit{et al}.~\cite{DBLP:journals/corr/abs-2312-06423} introduce MalPurifier, which aims to purify the adversarial perturbations within the adversarial malware sample to enhance the robustness of ML-based AMD systems.
Li \textit{et al}.~\cite{DBLP:journals/tdsc/LiCLXXX24} develop the PAD framework to offer convergence guarantees for robust optimization methods in the discrete feature space for adversarial training.

However, these methods do not fully address the practical requirements for defending against adversarial Android malware samples.
For instance, their defense effectiveness diminishes under adversarial malware attacks in the problem space~\cite{DBLP:conf/ccs/HeX0J23}.
Additionally, they may significantly negatively impact the original detection performance of ML-based AMD systems.

\section{Conclusion}

This paper introduces \system, a practical adversarial Android malware defense framework designed as a plug-in to enhance the adversarial robustness of ML-based AMD systems.
Based on the key observation that adversarial Android malware exhibits incompatibility between the perturbable feature space and the imperturbable feature space, \system employs encoder models with a customized contrastive learning loss to measure the incompatibility score, thereby defending against adversarial Android malware.
Our experimental results, conducted on large-scale real-world datasets, demonstrate that \system is effective against the SOTA adversarial Android malware attack methods across various target models.
Additionally, we evaluate the \system's performance in adaptive attack scenarios and with real-world antivirus solutions, finding that it also maintains strong defense effectiveness in these contexts.
The findings highlight the potential of \system as a valuable framework for defending against adversarial Android malware attacks.

\section*{Ethics considerations}

\label{sec:ethic}
The primary objective of our study is to improve the robustness of ML-based AMD systems, a topic with established precedence in earlier works~\cite{DBLP:conf/esorics/GrossePMBM17,DBLP:conf/sp/PierazziPCC20,DBLP:conf/ccs/HeX0J23}.
To defend against adversarial Android malware attacks, we generate adversarial Android malware in the problem space to evaluate the effectiveness of our defense method.
Despite the strict intent to defend against adversarial Android malware attacks, our research entails potential ethical concerns.

To minimize risks, we have taken several precautions.
First, we do not release the generated adversarial Android malware samples publicly.
Second, during our VirusTotal experiments, we utilized VirusTotal in the same manner as an ordinary user would by submitting applications for analysis.
Additionally, we contacted the VirusTotal team to share our findings and report our generated adversarial Android malware samples, providing their SHA256 hashes for further processing, such as removal or flagging.
While VirusTotal informed us on January 22, 2025, that their policy does not allow the removal of submitted samples, we have ensured that the SHA256 hashes of these adversarial malware samples have not been disclosed publicly, thus preventing easy reuse or retrieval by malicious actors.

We further limit potential misuse by restricting code and metadata (hashes) sharing to vetted research institutions and independent researchers upon request.
Verification will follow established precedents~\cite{DBLP:conf/sp/PierazziPCC20,DBLP:conf/ccs/ZhaoZZZZLYYL21,DBLP:conf/ccs/HeX0J23} and may include institutional affiliation and a signed agreement outlining responsible use.
During the study, all adversarial malware samples were handled in a controlled research environment, with appropriate security measures to prevent unauthorized access.

\section*{Open science}

To facilitate further research and the reproducibility of our experiments, the code and data of \system are responsibly shared with other researchers following previous works~\cite{DBLP:conf/sp/PierazziPCC20,DBLP:conf/ccs/ZhaoZZZZLYYL21,DBLP:conf/ccs/HeX0J23} due to potential ethical concerns.
To be specific, the code and data of \system will be open-sourced except for the part related to the adversarial Android malware generation.
The code and data related to the adversarial Android malware generation will be shared responsibly with vetted research institutions and independent researchers upon request. 

\bibliographystyle{plain}
\bibliography{reference}

\appendix

\section{Framework algorithm}
\label{appendix:algorithm}

The primary procedure of \system is shown in Algorithm~\ref{alg:overview}.
Within this algorithm, F, $g$ and $\mathbb{D}$ stand for the feature space, classifier and training set of the ML-based AMD method, respectively.
PS and IPS refer to the perturbable feature space and imperturbable feature space, respectively.
EPS and EIPS denote the encoder models used to project the perturbable and imperturbable features, respectively.
T signifies the threshold for compatibility score.
$z_{ps}$ and $z_{ips}$ represent the embeddings of the perturbable features and imperturbable features, respectively.
IncalScore symbolizes the incompatibility score of the test application sample.

\begin{algorithm}[t]
    \footnotesize
    \caption{\system}
    \label{alg:overview}
    \begin{algorithmic}[1]
        
        \Require{
        Feature space F;
        Target classifier $g$;
        Test application sample $z$;
        Training dataset $\mathbb{D}$;
        Calibration dataset $\mathbb{C}$;
        Quantification applications $\mathbb{Q}$;
        Feasible perturbation set $\mathbb{P}$.
        }
        
        \Ensure{
        Detection result $y$.
        }
        
        \Statex
        \State PS, IPS $ \leftarrow $ SpaceQuantification(F, $\mathbb{Q}$, $\mathbb{P}$) \Comment{Quantify the perturbable feature space and imperturbable feature space.}
        \State EPS, EIPS $ \leftarrow $ BuildEncoders(PS, IPS, $g$, $\mathbb{D}$) \Comment{Build the encoder models for feature projection.}
        \State T $ \leftarrow $ ThresholdCal(EPS, EIPS, $\mathbb{C}$) \Comment{Calibrate the threshold.}
        \State $y \leftarrow g(z)$ \Comment{Obtain the original detection result.}
        \If {$y$ is malicious}
            \State \Return $y$; \Comment{\system only revisits benign outputs.}
        \Else
            \State $z_{ps} \leftarrow$ EPS(PS($z$)) \Comment{Obtain and project the perturbable features.}
            \State $z_{ips} \leftarrow$ EIPS(IPS($z$)) \Comment{Obtain and project the imperturbable features.}
            \State IncalScore $ \leftarrow \ell_2(z_{ps}, z_{ips}) $ \Comment{Obtain the incompatibility score.}
            \If {IncalScore $ > T $} \Comment{Revisit the detection result.}
                \State $y \leftarrow $ malicious
            \Else
                \State $y \leftarrow $ benign
            \EndIf
        \EndIf
        \State \Return $y$;
        
    \end{algorithmic}
\end{algorithm}

\system begins by quantifying the perturbable feature space and imperturbable feature space of the ML-based AMD method using space quantification applications and the feasible perturbation set (line 1).
Next, \system trains the encoder models for projecting the perturbable features and imperturbable features, leveraging the training dataset and classifier of the ML-based AMD method (line 2).
Following this, \system calibrates the threshold of compatibility score using the calibration dataset (line 3).
For a test application, \system initially obtains the detection result from the original classifier (line 4).
If the test application is classified as malicious, \system retains this detection result (line 6); otherwise, \system revisits the sample for further analysis.
During the revisiting process, \system computes the embeddings of the perturbable features and imperturbable features for the test application sample (lines 8-9).
Subsequently, \system calculates the compatibility score for the test application sample (line 10).
If the compatibility score exceeds the threshold, \system classifies the sample as malicious; otherwise, it classifies the sample as benign (lines 11-15).

\section{Space Quantification Algorithm}
\label{appendix:spacequantify}

\begin{algorithm}[t]
    \footnotesize
    \caption{Space Quantification}
    \label{alg:spacequantify}
    \begin{algorithmic}[1]
        
        \Require{
        Feature space F;
        Quantification applications $\mathbb{Q}$;
        Feasible perturbation set $\mathbb{P}$.
        }
        
        \Ensure{
        Perturbable feature space PS;
        Imperturbable feature space IPS.
        }
        
        \Statex
        \State PS $ \leftarrow \emptyset $
        \For { $q$ in $\mathbb{Q}$ }
            \State $f_q \leftarrow $ F($q$) \Comment{Obtain the original feature.}
            \For { $P$ in $\mathbb{P}$ }
                \State $q' \leftarrow $ Implement($q$, $P$) \Comment{Obtain the original feature.}
                \State $f_{q'} \leftarrow $ F($q'$) \Comment{Obtain the perturbed feature.}
                \State PS $ \leftarrow $ PS $\cup$ Diff($f_q$, $f_{q'}$) \Comment{Identify the perturbable features.}
            \EndFor
        \EndFor
        \State IPS $ \leftarrow $ F $\setminus$ PS \Comment{Obtain the imperturbable feature space.}
        \State \Return PS, IPS;
        
    \end{algorithmic}
\end{algorithm}

The details of the space quantification algorithm can be found in Algorithm~\ref{alg:spacequantify}.
For each space quantification application, the space quantification algorithm first obtains its original feature vector (line 3).
Then, the space quantification algorithm iteratively implements the perturbation from the feasible perturbation set and observes the changes in features, identifying the altered features as perturbable features (lines 4-8).
Finally, the remaining features are classified as imperturbable features.

\section{Exponential dimension reduction policy}
\label{appendix:dimension}

\begin{algorithm}[t]
    \footnotesize
    \caption{Exponential Dimension Reduction Policy}
    \label{alg:dimension}
    \begin{algorithmic}[1]
        
        \Require{
        Output layer dimension $d$;
        Weight $w$;
        Max hidden dimension $M$;
        }
        
        \Ensure{
        Hidden dimension list $\mathbb{H}$.
        }
        
        \Statex
        \State $\mathbb{H} \leftarrow $ Empty List
        \State $h \leftarrow d * w$ \Comment{The current hidden dimension.}
        \While {True}
            \State $\mathbb{H}$.append($h$) \Comment{Add the current hidden dimension.}
            \State $h \leftarrow d * w$
            \If { $h \geq M$}   \Comment{Terminate.}
                \State \textbf{break}
            \EndIf
        \EndWhile
        \State $\mathbb{H}$.reverse()
        \State \Return $\mathbb{H}$;
        
    \end{algorithmic}
\end{algorithm}

As depicted in Algorithm~\ref{alg:dimension}, the exponential dimension reduction policy begins with the empty list (line 1).
The policy then iteratively adds dimensions by multiplying the exponential weight with the output layer dimension (lines 2-4).
This process continues, increasing the hidden dimension according to the exponential weight until the hidden dimension exceeds the maximum size (lines 5-7).

\section{Pseudo adversarial malware generation}
\label{appendix:pseudo}

\begin{algorithm}[t]
    \footnotesize
    \caption{Pseudo Adversarial Malware Generation}
    \label{alg:pseudo}
    \begin{algorithmic}[1]
        
        \Require{
        Malicious sample set $\mathbb{D}_m$;
        ML classifier $g$;
        Attempt budget $N$;
        Perturbable feature space PS;
        }
        
        \Ensure{
        Pseudo adversarial malware set $\mathbb{D}_{pam}$.
        }
        
        \Statex
        \State $\mathbb{D}_{pam} \leftarrow \emptyset$
        \For { $x_m$ in $\mathbb{D}_m$ }
            \State $n \leftarrow 0$ \Comment{$n$ represents the attempt times.}
            \While{ $n < N$ }
                \State $x_{pam} \leftarrow$ RandomPerturb($x_m$, PS)    \Comment{Random perturb.}
                \State $y \leftarrow g(x_{pam})$    \Comment{Obtain the model feedback.}
                \If {$y$ is benign}
                    \State $\mathbb{D}_{pam}$.add($x_{pam}$)    \Comment{Generation successful.}
                    \State \textbf{break}
                \EndIf
                \State $n \leftarrow n + 1$
            \EndWhile
        \EndFor
        \State \Return $\mathbb{D}_{pam}$;
        
    \end{algorithmic}
\end{algorithm}

The details of the pseudo adversarial malware generation algorithm can be found in Algorithm~\ref{alg:pseudo}.
To generate the set of pseudo adversarial malware samples, the algorithm iteratively perturbs each malicious sample within the malicious sample set (line 2).
For each malicious sample, the algorithm randomly perturbs features within the perturbable feature space and then obtains model feedback on the perturbed feature vector (lines 5-6).
If the perturbed feature vector is classified as benign, the pseudo adversarial malware generation process for that malicious sample is deemed successful (lines 7-10).

\section{Details of target model}
\label{appendix:targetmodel}

\paragraphbe{Implementation}
For the Drebin and SVM classifier, we utilize the same implementation provided by Pierazzi \textit{et al}.~\cite{DBLP:conf/sp/PierazziPCC20}.
The MLP classifier is implemented using PyTorch, following the methodology described by Grosse \textit{et al}.~\cite{DBLP:conf/esorics/GrossePMBM17}.
As for APIGraph, we utilize its official code available at \url{https://github.com/seclab-fudan/APIGraph}.
Regarding MaMadroid, we follow the implementation provided by He \textit{et al}.~\cite{DBLP:conf/ccs/HeX0J23} using Androguard \cite{Androguard}.
The RF classifier is implemented using scikit-learn, with the number of trees set to 50.

\begin{table}[t]\centering
	\caption{Detection performance of the target ML-based AMD systems.}
 
	\begin{tabular}[centering,width=0.5*\linewidth]{@{}C{1.3cm}C{1.6cm}C{1.2cm}C{0.8cm}C{0.5cm}C{1.1cm}@{}}
		\toprule
            Dataset Split & AMD Method & Precision & Recall & F1 & AUROC\\
		\midrule
        \multirow{4}{*}{Time} & Drebin & 0.84 & 0.82 & 0.83 & 0.95 \\
         & Drebin-DL & 0.87 & 0.81 & 0.84 & 0.96 \\
         & APIGraph & 0.84 & 0.79 & 0.82 & 0.95 \\
         & MaMadroid & 0.95 & 0.83 & 0.88 & 0.95 \\
         \midrule
        \multirow{4}{*}{Random} & Drebin & 0.91 & 0.88 & 0.89 & 0.98 \\
         & Drebin-DL & 0.89 & 0.85 & 0.87 & 0.98 \\
         & APIGraph & 0.90 & 0.88 & 0.89 & 0.98 \\
         & MaMadroid & 0.95 & 0.84 & 0.89 & 0.98 \\ 
        \bottomrule
  	\end{tabular}
 
	\label{tab:AMDPerformance}
\end{table}

\paragraphbe{Detection performance}
The detection performance metrics, including precision, recall, F1 score, and AUROC for the target ML-based AMD systems, are detailed in Table~\ref{tab:AMDPerformance}.
The true positive rates (TPRs) of the target ML-based AMD systems exceed 0.80 in most cases, indicating their effectiveness in malware detection.
Furthermore, models trained using the time-aware split method exhibit lower detection effectiveness compared to those trained using the random split due to the concept drift problem.

\section{Details of target attack methods}
\label{appendix:attackmethod}

\begin{table}[t]\centering
    \setlength{\abovecaptionskip}{0pt}
	\caption{The attack performance of the target three adversarial Android malware attack methods against the target ML-based AMD systems measured by ASR.}
 
	\begin{tabular}[centering,width=0.5*\linewidth]{@{}C{1.2cm}C{1.6cm}C{1.9cm}C{1.1cm}C{1.0cm}@{}}
		\toprule
            Dataset Split & AMD Method & AdvDroidZero & MAB & RA \\
		\midrule
        \multirow{4}{*}{Time} & Drebin & 63.77\% & 52.49\% & 60.92\% \\
         & Drebin-DL & 71.25\% & 62.08\% & 63.04\% \\
         & APIGraph & 73.37\% & 58.25\% & 63.74\% \\
         & MaMadroid & 93.50\% & 93.01\% & 90.98\% \\ 
         \midrule
        \multirow{4}{*}{Random} & Drebin & 67.36\% & 43.68\% & 56.20\% \\
         & Drebin-DL & 65.05\% & 50.73\% & 52.54\% \\
         & APIGraph & 68.05\% & 58.99\% & 64.16\% \\
         & MaMadroid & 95.40\% & 91.02\% & 91.72\% \\ 
        \bottomrule
  	\end{tabular}
 
	\label{tab:AttackRes}
\end{table}

\begin{table}[t]\centering
    \setlength{\abovecaptionskip}{0pt}
	\caption{The number of adversarial Android malware successfully generated by the three target adversarial Android malware attack methods, capable of evading the ML-based AMD system, selected from a random sample of 1,000 true positive malicious samples.}
 
	\begin{tabular}[centering,width=0.5*\linewidth]{@{}C{1.3cm}C{1.6cm}C{1.9cm}C{1.0cm}C{1.0cm}@{}}
		\toprule
            Dataset Split & AMD Method & AdvDroidZero & MAB & RA \\
		\midrule
        \multirow{4}{*}{Time} & Drebin & 433 & 316 & 380 \\
         & Drebin-DL & 451 & 388 & 411 \\
         & APIGraph & 474 & 396 & 429 \\
         & MaMadroid & 748 & 759 & 736 \\ 
         \midrule
        \multirow{4}{*}{Random} & Drebin & 578 & 363 & 462 \\
         & Drebin-DL & 549 & 417 & 434 \\
         & APIGraph & 621 & 512 & 429 \\
         & MaMadroid & 891 & 851 & 864 \\ 
        \bottomrule
  	\end{tabular}
 
	\label{tab:AttackResAct}
\end{table}

The attack success rate for each case is presented in Table~\ref{tab:AttackRes}, demonstrating the effectiveness of the three adversarial Android malware attack methods against the target ML-based AMD systems.
Table~\ref{tab:AttackResAct} provides the actual number of adversarial Android malware samples successfully generated by the three target adversarial Android malware attack methods.
The three adversarial Android malware attack methods can generate sufficient adversarial Android malware samples for evaluation.
Recognizing the bugs and corner cases present in the FlowDroid~\cite{DBLP:conf/pldi/ArztRFBBKTOM14}, as discussed by previous works~\cite{DBLP:conf/sp/PierazziPCC20,DBLP:conf/ccs/HeX0J23}, we also encounter crashes during the modification of APKs with FlowDroid.
In accordance with previous works~\cite{DBLP:conf/sp/PierazziPCC20,DBLP:conf/ccs/HeX0J23}, these crashes are excluded from the attack success rate computation.

\section{Average defense performance of thresholds}
\label{appendix:avg_defense}

\begin{table*}[t]\centering
	\caption{The average defense performance of all thresholds against AdvDroidZero across different target models measured by the three metrics. TI@K represents the TNIR at K\% in the calibration dataset.}
 
	\begin{tabular}[centering,width=0.5*\linewidth]{@{}C{1.2cm}C{1.6cm}C{1.3cm}C{1.0cm}C{1.0cm}C{1.0cm}C{1.0cm}C{1.0cm}C{1.0cm}C{1.0cm}C{1.0cm}C{1.0cm}@{}}
		\toprule
            \multirow{2}{*}{\begin{tabular}[c]{@{}c@{}}	Dataset \\ Split \end{tabular}} & \multirow{2}{*}{\begin{tabular}[c]{@{}c@{}}	AMD \\ Method \end{tabular}} & \multirow{2}{*}{\begin{tabular}[c]{@{}c@{}}	Classifier \end{tabular}} & \multicolumn{3}{c}{NDASR} & \multicolumn{3}{c}{TNIR} & \multicolumn{3}{c}{FNIR} \\ \cmidrule(l){4-6} \cmidrule(l){7-9} \cmidrule(l){10-12}
             & & & TI@10 & TI@5 & TI@1 & TI@10 & TI@5 & TI@1 & TI@10 & TI@5 & TI@1 \\
		\midrule
        \multirow{5}{*}{Time} & Drebin & SVM & 85.84\% & 84.95\% & 55.44\% & 2.94\% & 2.73\% & 0.30\% & 54.58\% & 54.18\% & 37.43\% \\
        \cmidrule(l){4-6} \cmidrule(l){7-9} \cmidrule(l){10-12}
         & Drebin-DL & MLP & 83.53\% & 83.28\% & 51.51\% & 2.91\% & 2.73\% & 0.17\% & 40.37\% & 40.18\% & 20.65\% \\
         \cmidrule(l){4-6} \cmidrule(l){7-9} \cmidrule(l){10-12}
         & APIGraph & SVM & 80.58\% & 80.43\% & 45.51\% & 2.77\% & 2.65\% & 0.16\% & 49.80\% & 49.68\% & 34.03\% \\
         \cmidrule(l){4-6} \cmidrule(l){7-9} \cmidrule(l){10-12}
         & MaMadroid & RF & 86.82\% & 86.08\% & 60.87\% & 3.78\% & 3.57\% & 1.03\% & 35.99\% & 35.53\% & 24.47\% \\ \midrule
        \multirow{5}{*}{Random} & Drebin & SVM & 91.84\% & 91.77\% & 87.28\% & 1.83\% & 1.70\% & 0.92\% & 50.24\% & 49.96\% & 38.68\% \\
        \cmidrule(l){4-6} \cmidrule(l){7-9} \cmidrule(l){10-12}
         & Drebin-DL & MLP & 89.58\% & 89.55\% & 88.23\% & 1.23\% & 1.14\% & 0.78\% & 52.61\% & 52.53\% & 50.58\% \\
         \cmidrule(l){4-6} \cmidrule(l){7-9} \cmidrule(l){10-12}
         & APIGraph & SVM & 92.67\% & 92.63\% & 89.63\% & 1.65\% & 1.55\% & 0.83\% & 46.28\% & 46.06\% & 36.01\% \\
         \cmidrule(l){4-6} \cmidrule(l){7-9} \cmidrule(l){10-12}
         & MaMadroid & RF & 95.26\% & 95.23\% & 92.03\% & 2.31\% & 2.18\% & 1.35\% & 67.82\% & 67.61\% & 62.66\% \\
        \bottomrule
  	\end{tabular}
 
	\label{tab:avgthreshold}
\end{table*}

As shown in Table~\ref{tab:avgthreshold}, the detection performance achieved by our threshold calibration algorithm surpasses the average detection performance in most cases.
For example, with the Drebin model using the SVM classifier trained on the time-aware split dataset under TNIR@1, the threshold calibration algorithm achieves an NDASR of 80.49\%, whereas the average detection performance is only NDASR of 55.44\% under the same setting.

\end{document}